\documentstyle[aps,pra]{revtex}
\begin{document}

\title{A unified approach for the solution of
the Fokker-Planck equation}
           
\author{G. W. Wei}
\address{Department of Computational Science, National 
         University of Singapore, Singapore 117543} 

\date{\today}
\maketitle
\begin{abstract}

This paper explores the use of a discrete 
singular convolution algorithm 
 as a unified approach
for numerical integration of the  Fokker-Planck equation.
The unified features of the discrete 
singular convolution algorithm are discussed.
It is demonstrated that different implementations of the 
present algorithm, such as global, local,
Galerkin, collocation, and finite difference, 
can be deduced from a single starting point.
Three benchmark stochastic systems, the repulsive 
Wong process, the Black-Scholes equation
and a genuine nonlinear model, are employed 
to illustrate the robustness and to test accuracy of 
the present approach for the solution of 
the Fokker-Planck equation via a time-dependent method. 
An additional example, the incompressible Euler equation,
is used to further validate the present 
approach for more difficult problems.
 Numerical results 
indicate that the present unified approach 
is robust and accurate 
for solving the Fokker-Planck equation.

\end{abstract}
\newpage

\section{Introduction}

Much research has been done in the 
exploration of accurate and stable computational 
methods for the numerical solution of the Fokker-Planck 
equation\cite{Suzuki,Risken,CCR,LarKos,IVMA,WehWol,BSS,Shizgal,BlaShi,KomShi,BCM,Haken2,GraGre,Dekker,FriNow,MunGar,Kho,ErmBuc,ForWas,PSMT,ChaCoo,LLPS,DroMor,Epperlein,weijcp}.
A detailed comparison of several different 
approaches was given by Park and Petrosian\cite{ParPet}
(see Ref. \cite{weijcp} for a literature review). 
   In fact, the solution of the Fokker-Planck equation,  
in particular the nonlinear form of this equation, is still 
a non-trivial problem. In somewhat a broader sense, 
finding numerical solutions for partial differential 
equations (PDEs) is still a challenge
owing to the presence of possible 
singularities and/or homoclinic manifolds
that induce sharp transitions in the solutions\cite{Kadanoff}.
These phenomena can be observed 
in many real systems such as black holes in astronomy, 
shock waves in compressible fluid flow, vortex sheets in 
incompressible flow associated with a high Reynolds number, 
and burst events in the turbulent boundary layer.
The difficulties associated with these phenomena can often be 
characterized by sharp changes occurring 
in a very small spatial region  which
can strongly influence the global properties of the system. 
The presence of these phenomena 
can be extremely sensitive to numerical algorithms and can easily 
lead to numerically induced spatial and/or temporal chaos\cite{Ablowitz}.  
At present, there are two major classes of numerical methods for 
solving PDEs, namely, global methods and local methods.
In global methods, unknown functions and their
derivatives are expanded in terms of a finite basis 
set with each element having a global support.  
The expansion coefficients are often determined by 
the method of tau, or Galerkin, or collocation, or others.
In the Tau method, the residual for a truncated expansion is 
required to be orthogonal to a subset of basis functions used 
in the expansion, which, together with the boundary conditions,
determines the expansion coefficients. In the global Galerkin method, 
a new set of basis functions is constructed by the superposition
of the original basis functions. The requirement of 
the residual be orthogonal to the new set of basis functions,
together with the boundary conditions, determines the 
expansion coefficients. 
In the global collocation approach, the residual vanishes
at a subset of node points of the highest order basis function
used in the expansion. The global collocation is also called
pseudospectral method.
Three most important local approaches are
finite difference, finite volume and finite element methods. 
In finite difference methods, the solution 
is interpolated in terms of a set of grid values; 
the spatial derivatives are usually approximated by 
algebraic expressions involving nearest 
neighbor grid points. 
In finite volume approaches, 
the emphasis is on a set of integro-differential equations
and their associated surface and volume integrations.
The values on the boundary of each ``numerical molecule'' 
are usually interpolated by low order schemes. 
The spatial derivatives
are approximated in the same way as those used in the 
finite difference methods. 
Finite element methods form one of the most versatile 
classes of numerical methods.
Depending on the system under study, finite element methods
can be formulated either in terms of the method of weight 
residuals or in terms of variational principles. 
Usually, PDEs are integrated by 
using a set of trial functions, each with a small 
region of support. The solution is represented by linear 
superpositions of these trial functions.

Global methods are highly localized in their spectral space, 
but are unlocalized in the coordinate space.  
By contrast, local methods have high spatial 
localization, but are delocalized in their spectral space.
In general, global methods are much more accurate than 
local methods, while the major advantage of local methods 
is their flexibility for handling complex geometries 
and boundary conditions. 
Moreover, the use of global methods is usually 
restricted to structured grids,
whereas, local methods can be implemented to block-structured 
grids and even unstructured grids.

There were hectic debates among the numerical computation communities
over the advantages and disadvantages of various
numerical methods in the past a few decades. 
These debates stimulated  
 the development of powerful numerical 
methods for a wide variety of science and engineering 
applications. Such development has, in association 
with the availability of 
inexpensive high-performance computers, 
led to the establishment of numerical simulations as 
an alternative approach for researches and applications. 
The connection of various numerical methods has always
been an important research topic. Finlayson discussed
the relation between the Galerkin and 
the Ritz variational principle\cite{Finlayson}.
Canuto et al rearranged their spectral basis functions so that 
some global collocation method can be regarded as a 
special case of certain global Galerkin 
methods\cite{CHQZ}. Fornberg addressed the 
common feature between pseudospectral methods and 
high order finite difference methods\cite{Fornberg2}. 
The connection between global and local methods can 
also be realized in the framework of the 
method of weighted residual by choosing trial functions of either
piecewise Lagrange polynomials or global Lagrange
polynomials. The connection of methods of 
finite element, finite difference and finite volume
is now well understood\cite{Dow}. 
However, to our knowledge, none has reported a unified
scheme for the discussion of all of 
the abovementioned methods.  

In  previous work\cite{weijcp}, we proposed a discrete
singular convolution (DSC) algorithm and demonstrated its 
use for the numerical solution of Fokker-Planck equation via  
eigenfunction expansions. 
The DSC algorithm was shown to be a potential numerical approach for 
Hilbert transform, Abel transform, Radon transform and delta transform.
Three standard problems, the Lorentz 
Fokker-Planck equation, the bistable model and the
Henon-Heiles system,  were utilized to test the
accuracy, reliability, and speed of convergence of 
the DSC-eigenfunction approach. 
All results were in excellent agreement 
with those of previous methods in the field. 
Recently, the DSC algorithm has been successfully tested 
for integrating the sine-Gordon equation with 
initial values close to homoclinic orbits\cite{weiphysca},
which is extremely difficult to compute because of 
the possible presence of numerical chaos\cite{Ablowitz}.
Excellent results are obtained for solving the 
Navier-Stokes equation and for engineering structural 
analysis\cite{weicmame}. 
The purpose of the present paper 
is twofold. First, we study the unified 
features of the DSC algorithm for treating
partial differential equations. This 
is accomplished by focusing on the DSC kernels of 
the delta type and their approximations. 
Second, we explore the use of the DSC as a 
unified approach for solving the Fokker-Planck 
equation via direct explicit time propagations.
The eigenfunction expansion approach 
provides a Schr\"{o}dinger-equation type picture for the
understanding of the Fokker-Planck equation. However, its 
use is restricted to a certain class of Fokker-Planck 
operators (essentially for the Fokker-Planck 
operators their equivalent Schr\"{o}dinger potentials are 
bounded from below). 
The present direct approach is 
applicable to a wider class of problems.
These two DSC-based  approaches
have the same level of accuracy for the 
numerical solution of the Fokker-Planck equation. 
They are complementary to each other for 
solving a wide variety of Fokker-Planck 
systems arising from practical situations.

This paper is organized as the follow. 
The unified features of the DSC algorithm
are discussed in Section II.    
We demonstrate that, the present DSC algorithm
provides a unified framework for solving 
the Fokker-Planck equation, and 
partial differential equations in general.
In particular, we show that 
various different implementations
of the DSC algorithm, 
such as global, local, Galerkin, collocation, 
and finite difference, can be deduced from a 
single starting point.
The application of the present DSC
approach to the solution of 
the Fokker-Planck equation and Euler equation 
is presented in Section III. 
We use four examples to illustrate the present approach.
The first example  is the repulsive Wong process which is 
useful for testing the ability of 
handling monomodality-bimodality transition.
The second example is the Black-Scholes equation for 
option derivatives. This is interesting stochastic model
for option pricing in financial market.
The third case  treated is a nonlinear stochastic 
model which has certain connection to a mean-field model 
for self-organization processes in biological systems such as
muscle contraction.
Notably, all of these problems are treated by an explicit
time-propagation approach in contrast to the 
eigenfunction expansion used in our previous work\cite{weijcp}.
Since the abovementioned examples are of strong 
parabolic type, we consider an additional problem, 
the incompressible Euler equation, to 
further validate the DSC approach for more difficult 
problems. The incompressible Euler equation is chosen 
because its equations for velocity vector and pressure field 
are of strong hyperbolic type and elliptic type, respectively. 
Thus, this last example is complimentary
to other three examples from the point of view of 
numerical analysis. This paper ends with a discussion.

\section{Properties of the discrete singular convolution}

This section presents the properties of the discrete 
singular convolution (DSC) algorithm for solving 
differential equations. The first subsection addresses 
the unified features of the DSC algorithm in the line
of the method of weighted residuals. 
 Relevant properties of DSC trial functions are discussed 
the second subsection.

\subsection{Unified features}

Without the loss of generality, 
it is  assumed that at a fixed time, 
a stochastic process 
is governed by a differential 
equation. To solve the differential 
equation, 
one can start with either by 
approximating the  original differential 
operator or by approximating 
the actual solution of the differential 
equation  while maintaining 
the original differential operator. 
The latter is accomplished by 
explicitly defining a functional form for 
approximations.
Let us assume that the differential 
equation has the form
\begin{equation}\label{eq1}
  {\cal  L} u({ x})  = f({ x}),   ~~ { x}\in \Omega,
\end{equation}
where ${\cal  L}$ is a linear operator and 
$u({ x})$ is the unknown solution of interest. Here 
$f(x)$ is a known force term, 
$\Omega$ denotes the domain over which the 
differential equation applies.

       The approximate solution is sought from a finite 
set of $N$ DSC trial functions of a given 
{\it resolution} $\alpha$, 
denoted by $S^{N,M}_{\alpha,\sigma}$ with $M$ being 
the half width of support of each element. 
Here $\sigma$ is a {\it regularization} parameter
for improving the regularity of the set.  
The case of regularization free is easily obtained by setting 
$\sigma\rightarrow\infty$.
Elements of the set $S^{N,M}_{\alpha,\sigma}$ can be explicitly 
given by $\{\phi^M_{\alpha,\sigma;1}, \phi^M_{\alpha,\sigma;2},...,
\phi^M_{\alpha,\sigma; N}\}$. For a given computational 
domain, the resolution parameter $\alpha$ is determined 
by $N$.

    An important property of the DSC trial functions 
$\{\phi^M_{\alpha,\sigma;k}\}$
is that when the  trial function is free of regularization,
each  member of the set is a {\it reproducing kernel} 
at highest resolution
\begin{equation}\label{prop1}
\lim_{\alpha\rightarrow\infty}
<\phi^M_{\alpha,\sigma; k},\eta >=\eta({ x}_k),
\end{equation}
where $< \cdot,\cdot >$ denotes the standard inner product.
In fact, if an appropriate 
basis is used for $\phi$ and the limit on $\sigma$ is
taken, $\phi$ of each resolution 
can be a reproducing kernel for $L^2$ functions 
bandlimited to appropriate sense. In general, we require
the {\it low pass filter property}
that for given $\alpha\neq0,\sigma\neq0$ and $M\gg 0$ 
\begin{equation}\label{prop2}
<\phi^M_{\alpha,\sigma; k},\eta >\approx\eta({ x}_k).
\end{equation}
This  converges uniformly
when the resolution is refined, e.g.,
$\alpha\rightarrow\infty$. 
Many examples of such DSC trial functions are given in 
Refs. \cite{weijcp} and \cite{weicpl}.
Further discussion on these functions is given in the next 
subsection.
Equations (\ref{prop1}) and (\ref{prop2}) are special 
requirements satisfied by the DSC kernels of delta 
type\cite{weijcp}.

    In the present DSC approach, an approximation 
to the function of interest $u(x)$ can be expressed 
as a linear combination   
\begin{equation}\label{eq2-3}
U^{N,M}_{\alpha,\sigma}({ x}) =\sum_{k=1}^{N}
U_{\alpha,\sigma; k} 
\phi^M_{\alpha,\sigma; k}({ x}),
\end{equation}
where ${x}$ is an independent variable and
$U_{\alpha,\sigma; k}$ 
is a DSC approximation to the solution wanted at point $x_k$.
This structure is due to 
the DSC trial function property (\ref{prop2}) and it
dramatically 
simplifies the solution procedure in
practical computations.

In this formulation, we choose the set 
$S^{N,M}_{\alpha,\sigma}$ a priori, 
and then determine the 
coefficients $\{ U_{\alpha,\sigma; k} 
\}$ 
so that $U^{N,M}_{\alpha,\sigma}({ x})$ is a good 
approximation to $u({ x})$. 
To determine $U_{\alpha,\sigma; k}$, we minimize the 
amount by which $U^{N,M}_{\alpha,\sigma}({ x})$ 
fails to satisfy the original governing 
equation (\ref{eq1}). A measure of 
this failure can be  defined as
\begin{equation}\label{eq3}
R^{N,M}_{\alpha,\sigma}({ x}) \equiv  
{\cal L} U^{N,M}_{\alpha,\sigma}({ x}) - f({ x}),
\end{equation}
where $R^{N,M}_{\alpha,\sigma}({ x})$ is the residual 
for particular choices of resolution, regularization and
half width of the support. Note that Eq. (\ref{eq3})
is similar to the usual statement in the method of 
weighted residuals. However,  the approximation 
$U^{N,M}_{\alpha,\sigma}({ x})$ is 
constructed by using the DSC trial functions,
$\phi^M_{\alpha,\sigma; k}({ x})$, in the present 
treatment. Let Eq. (\ref{eq1}) and its associated 
boundary conditions be well-posed, 
then there exists 
a unique solution $u({ x})$ which generally resides 
in an infinite-dimensional space. 
Since the DSC approximation $U^{N,M}_{\alpha,\sigma}$ is 
constructed from a finite-dimensional set, 
it is generally the case 
that $U^{N,M}_{\alpha,\sigma}({ x})\neq u({ x})$ and 
therefore $R^{N,M}_{\alpha,\sigma}({ x}) \neq 0$.

{\it Galerkin.}~
We seek to optimize $R^{N,M}_{\alpha,\sigma}({ x})$
 by forcing it to zero in 
a {\it weighted average sense} over the domain 
$\Omega$. A convenient starting point is 
the Galerkin 
\begin{equation}\label{eq4}
\int_\Omega R^{N,M}_{\alpha,\sigma}({ x}) 
\phi^{M'}_{\alpha',\sigma'; l} ({ x}) d{ x} = 0,~~
\phi^{M'}_{\alpha',\sigma'; l}({ x}) 
\in S^{N',M'}_{\alpha',\sigma'}, 
\end{equation}
where the  weight set 
$S^{N',M'}_{\alpha',\sigma'}$  can be simply 
chosen being identical to the DSC trial function set
$S^{N,M}_{\alpha,\sigma}$. We refer
Eq. (\ref{eq4}) as a DSC-Galerkin statement.

{\it Collocation.}~
First, we note that in view of Eq. (\ref{prop1}), 
the present DSC-Galerkin statement reduces to 
a collocation one at the limit of $\alpha'$
\begin{equation}\label{eq5}
\lim_{\alpha'\rightarrow\infty}
\int_\Omega R^{N,M}_{\alpha,\sigma}({ x}) 
\phi^{M'}_{\alpha',\sigma'; l} ({ x}) d{ x}=
R^{N,M}_{\alpha,\sigma}({ x}_l)=0,
\end{equation}     
where $\{{  x}_l \}$ is the set of collocation points.
However, in digital computations, we cannot take the above
limits. It follows from the low pass filter property of the 
DSC trial functions, Eq. (\ref{prop2}), that 
\begin{equation}\label{eq6}
\int_\Omega R^{N,M}_{\alpha,\sigma}({ x}) 
\phi^{M'}_{\alpha',\sigma'; l} ({ x}) d{ x}\approx
R^{N,M}_{\alpha,\sigma}({ x}_l)\approx0.
\end{equation}     
It can be proven that for appropriate choice of 
$S^{N',M'}_{\alpha',\sigma'}$,
the first approximation of Eq. (\ref{eq6}) 
converges uniformly. 
The difference between
the true DSC-collocation, 
\begin{equation}\label{eq5-2}
\lim_{\alpha'\rightarrow\infty}
R^{N,M}_{\alpha,\sigma}({ x}_l)=0,
\end{equation}     
and the {\it Galerkin induced collocation}, (\ref{eq6}), 
diminishes to zero for 
appropriate DSC trial functions.

{\it Global and local.}~
Global approximations to a function and its derivatives are realized
typically by a set of truncated $L^2(a,b)$ function expansions.
It is called global because the values of a function and its
derivatives at a particular point $x_i$ in the coordinate
space involve the {\it full} set of grid points in a computational 
domain $\Omega$. Whereas a local method does so by requiring only 
a few nearest neighbor points.
In the present DSC approach, since the choices of $M$ and/or $M'$ are 
independent of $N$, one can choose $M$ and/or $M'$
so that a function and its
derivatives at a particular point $x_l$ are approximated either
by the full set of grid points in the computational 
domain $\Omega$ or just by a few nearest neighbor grid points.
In fact, this freedom for the selection of $M$ endows the DSC 
algorithm  with {\it controllable accuracy} for solving 
differential equations and the flexibility for handling 
complex geometries.

{\it Finite Difference.}~
In the finite difference method, the differential 
operator is approximated by difference operations.
In the present approach, the DSC-collocation 
expression of Eq. (\ref{eq6})
is equivalent to a $2M+1$ (or $2M$) term  finite difference method.
This follows from the fact that the DSC approximation to the 
$n$th order derivative of a function can be rewritten as 
\begin{equation}\label{eq10}
\left.{d^q u\over dx^q}\right|_{x=x_k}
\approx
\sum_{l=k-M}^{k+M}c^{q}_{kl,M}u(x_l),
\end{equation}     
where $c^{q}_{kl,M}$ are a set of DSC weights for the finite
difference approximation and are given by 
\begin{equation}\label{eq11}
c^{q}_{kl,M}= 
\left.{d^q \over dx^q}\phi^{M}_{\alpha,\sigma; l} ({ x})
\right|_{{ x}={ x}_k}. 
\end{equation}     
Obviously, for each different choice of 
$\phi^M_{\alpha,\sigma}$, we have a different 
DSC-finite difference approximation. 
Hence, the present DSC approach is a generalized finite 
difference method. This DSC-finite difference was tested 
in previous studies\cite{weicpl}.
When $M=1$, the DSC-finite difference
approximation reaches its low order 
limit and the resulting matrix is tridiagonal.
In this case,
the present DSC weights $c^{q}_{kl,M}$
can always be made exactly the same as those of  
the second order central difference scheme 
(i.e. ${1\over2\Delta}, 0, -{1\over2\Delta}$ for the first order 
derivative and ${1\over\Delta^2}, -{2\over\Delta^2}, {1\over\Delta^2}$
for the second order derivative. Here
$\Delta$ is the grid spacing.) of the 
standard finite difference method
by appropriately choosing the parameter $\sigma$.
However, even in this case, the
DSC-finite difference approximation
does not have to be the same as the standard 
finite difference scheme 
and can be optimized in a practical 
application by varying $\sigma$.

\subsection{DSC trial functions}
  
There are many DSC trial functions
that satisfy Eq. (\ref{prop2}). The requirement 
of  Eq. (\ref{prop2}) can be regarded as an
{\it approximate reproducing kernel} or quasi reproducing 
kernel. The reason for using an approximate reproducing 
kernel can be understood from the following 
analysis of Shannon's kernel ${\sin(\alpha x)\over \pi x}$.
Shannon's kernel is a delta sequence 
\begin{eqnarray}\label{Shannon}    
\lim_{\alpha\rightarrow\infty} 
{\sin(\alpha x)\over \pi x}=\delta(x),           
\end{eqnarray}   
where $\delta(x)$ is the delta distribution which can be 
regarded as a {\it universal reproducing kernel}
because its Fourier transform is the unit.
However, such a universal reproducing kernel cannot be directly 
used in digital computations because it is a distribution 
(Precisely, it is belongs to Sobolev space of order -1, 
$H^{-1}$.) and it does not have a value anywhere in the 
coordinate space. Therefore, in certain sense, 
constructing a reproducing kernel in an appropriate
$L^2(a,b)$ space is equivalent to finding a 
sequence of approximation of the delta distribution in
the  $L^2(a,b)$.
In fact, Shannon's kernel is an element of the 
Paley-Wiener reproducing kernel Hilbert space 
$B^2_\pi$ 
\begin{equation}\label{eq80}  
f(x)=\int_{-\infty}^{\infty}f(y)  
{\sin\pi(x-y)\over\pi(x-y)}dy  
,~~~~\forall f\in B^2_\pi,    
\end{equation}  
where $\forall f\in B^2_\pi$ 
indicates that, in its Fourier representation, 
the $L^2$ function $f$ vanishes outside   
the interval $[-\pi,\pi]$.      
What is important for digital computations
is the fact that the Paley-Wiener reproducing 
kernel Hilbert space has a 
{\it sampling basis} $S_k(x)$  
\begin{equation}\label{eq60}  
S_k(x) =   
{\sin\pi(x-y_k)\over\pi(x-y_k)},~~y_k=k,~~\forall k\in {\cal Z},  
\end{equation}  
where symbol ${\cal Z}$ denotes the set of all integers.  
Expression (\ref{eq60}) 
 provides a discrete representation of every  
(continuous) function in ${B^2_\pi}$  
\begin{equation}\label{eq40}  
f(x)=\sum_{k\in {\cal Z}}f(y_k)S_k(x),~~~~\forall f\in {B^2_\pi}.  
\end{equation}  
This is  Shannon's sampling theorem and 
is particularly important to  
information theory and the theory of sampling.
Note that Shannon's kernel is obviously  
interpolative on ${\cal Z}$
\begin{equation}\label{eq50}  
S_n(x_m)=\delta_{n,m},  
\end{equation}  
where $\delta_{n,m}$ is the  Kronecker delta function.    
Computationally, being interpolative is of particular 
importance for numerical accuracy and simplicity.   
  
In wavelet analysis, ${\sin(\pi x)\over \pi x}$ is  
Shannon's wavelet scaling function 
and its Fourier transform 
is a characteristic function, i.e. it is an unsmoothed,  
{\it ideal} low pass filter. 
In physical language, it is a projection to the 
frequency subband $[-\pi, \pi]$.    
By the Heisenberg uncertainty principle, such a (sharp)
projection must be an infinite impulse response (IIR) filter.
The usefulness of such a filter is limited because 
it is de-localized in the coordinate space and
requires infinitely many sampling data.
In practical computations, a truncation is required, 
which leads to large truncation error and even worse, 
numerical instability.
To improve the smoothness and regularity of  
Shannon's kernel, we introduce a 
regularization  
\begin{equation}\label{rsinc4-2}    
{\Phi}_{\sigma}(x) = {\sin(\pi x)\over \pi x}R_{\sigma}(x)
~~~~(\sigma>0),    
\end{equation}     
where $R_{\sigma}$ is a {\it regularizer} which     
has properties    
\begin{equation}\label{r0}     
\lim_{\sigma\rightarrow\infty}R_{\sigma}(x)=1    
\end{equation}     
and    
\begin{equation}\label{r1}     
R_{\sigma}(0)=1.    
\end{equation}    
Here Eq. (\ref{r0}) is a general condition     
that a  regularizer must satisfy,     
while Eq. (\ref{r1}) is specifically for a {\it delta     
regularizer}, which is used in regularizing a     
delta kernel.    
Various delta regularizers 
 can be used for numerical  
computations. An excellent one is the Gaussian 
\begin{equation}\label{r2}     
R_{\sigma}(x)=\exp\left[{-{x^2\over2\sigma^2}}\right].    
\end{equation}     
An immediate benefit of the regularized Shannon's kernel,  
Eq. (\ref{rsinc4-2}), is that its Fourier transform     
is infinitely differentiable because the Gaussian is  
an element of the Schwartz class functions.   
Qualitatively, all kernels of the Dirichlet type 
oscillate in the coordinate representation. Specifically, 
Shannon's kernel has a long tail which is  
proportional to ${1\over x}$, whereas, 
the regularized kernels decay exponentially fast,  
especially when the $\sigma$ is very small. 
In the Fourier representation, 
regularized Shannon's kernels have an ``optimal''  
shape in their frequency responses. 
Of course, they all reduce to Shannon's low pass filter at  
the limit 
 \begin{equation}\label{rsinc1-2}    
\lim_{\sigma\rightarrow\infty}\Phi_\sigma(x)=     
\lim_{\sigma\rightarrow\infty}{\sin\pi x\over \pi x}    
e^{-{x^2\over 2\sigma^2}}={\sin\pi x\over \pi x}.  
\end{equation}    
Quantitatively, one can    
examine the normalization of $\Phi_\sigma(x)$       
\begin{eqnarray}\label{rsinc3}    
\int\Phi_{\sigma}( x )dx \nonumber 
&=&\hat{\Phi}_{\sigma}(0) \\ 
&=& \sqrt{2\pi}  {\sigma}    
\sum_{k=0}^{\infty}{(-1)^k\over k!(2k+1)}    
\left({\pi\sigma \over\sqrt{2} }\right)^{2k}\\
&=&
{\rm erf}\left({\pi\sigma\over\sqrt{2}}\right)\nonumber\\    
&=&1-\sqrt{2\over\pi}{1\over\sigma}e^{-{\sigma^2\pi^2\over2 }}    
\int_0^{\infty }e^{-{t^2\over2\sigma^2}-{\pi t}}dt\nonumber\\    
&=&1-{\rm erfc}\left({\pi\sigma\over\sqrt{2}}\right)\\   
&\neq&1,
\end{eqnarray} 
where ${\rm erf}(z)={2\over\sqrt{\pi}}\int_{0}^{z}e^{-t^2}dt$
is the error function and
erfc(z) is the complementary error function.    
Note that for a given  $\sigma>0$,      
erfc(${\pi\sigma\over\sqrt{2}}$) is positive definite.    
Thus, $\hat{\Phi}_{\sigma}(0)$     
is always less than unity except at the limit of     
$\sigma\rightarrow\infty$.  
Therefore, $\Phi_{\sigma}(x)$ 
is no longer a reproducing kernel.
However, we argue that $\Phi_{\sigma}(x)$ is an approximate 
reproducing kernel because 
when we choose $\sigma \gg \sqrt{2}/\pi$, which is     
the case in many practical applications, the residue term,     
erfc(${\pi\sigma\over\sqrt{2}}$), approaches     
zero very quickly.   As a result,  
$\hat{\Phi}_{\sigma}(0)$ is extremely close to unity. 
As trial functions, regularized Shannon's kernels
do not form a sampling basis. They are no longer 
orthogonal in general. However, they just {\it slightly} 
miss the orthogonality
and the requirement of a basis.

For numerical computations, it turns out that the approximate
reproducing kernel has much less truncation errors
for interpolation and numerical differentiations.
Qian and the present author\cite{qian}  have recently 
given the following theorem for truncation errors.\\
{\bf Theorem} Let $f$ be a function $f\in L^2(R)\cap C^{s}(R)$ 
and bandlimited to 
$B$, $(B< \frac{\pi}{\Delta},~ \Delta$ is the grid spacing).  
For a fixed $t\in R $ 
and $\sigma>0$, denote $g(x)=f(x)H_k(\frac{t-x}{\sqrt{2}\sigma})$, 
where $H_k(x)$ is the
$k$th order Hermite polynomial. If $g(x)$ satisfies 
\begin{equation} 
g'(x)\leq
g(x)\frac{(x-t)}{{\sigma}^2} 
\end{equation} 
for $x\geq t+(M_1-1)\Delta$, and
\begin{equation} 
g'(x)\geq g(x)\frac{(x-t)}{{\sigma}^2} 
\end{equation} 
for $x\leq
t-M_2\Delta$, where $M_1,M_2\in \mathcal{N}$,  then for any 
$s\in \mathcal{Z}^{+}$
\begin{eqnarray} 
&&\left\|
f^{(s)}(t)-\sum_{n=\lceil\frac{t}{\Delta}\rceil-M_2}^{\lceil
\frac{t}{\Delta}
\rceil+M_1}f(n\Delta)
\left[\frac{\sin\frac{\pi}{\Delta}(t-n\Delta)}{\frac{\pi}{\Delta}
(t-n\Delta)}
\exp(-\frac{(t-n\Delta)^2}{2{\sigma}^2})\right]^{(s)}
\right\|_{L^2(R)}\nonumber\\ 
&&\leq
\sqrt{3}\left[\frac{\| f^{(s)}(t)\|_{L^2(R)}}{2\pi\sigma
(\frac{\pi}{\Delta}-B)
\exp(\frac{{\sigma}^2(\frac{\pi}{\Delta}-B)^2}{2})}\right.
\nonumber\\ 
&&\left.+ \frac{\|
f(t)\|_{L^2(R)}\sum_{i+j+k=s}\frac{s!{\pi}^{i-1}
H_k(\frac{-M_1\Delta}{\sqrt{2}\sigma})}
{i!k!{\Delta}^{i-1}(\sqrt{2}\sigma)^k((M_1-1)\Delta)^{j+1}}}
{\exp(\frac{(M_1\Delta)^2}{2{\sigma}^2})} \right.\nonumber\\ 
&&\left.+\frac{\|
f(t)\|_{L^2(R)}\sum_{i+j+k=s}\frac{s!{\pi}^{i-1}
H_k(\frac{-M_2\Delta}{\sqrt{2}\sigma})}
{i!k!{\Delta}^{i-1}(\sqrt{2}\sigma)^k(M_2\Delta)^{j+1}}}
{\exp(\frac{(M_2\Delta)^2}{2{\sigma}^2})}\right],
\label{eq30} 
\end{eqnarray}
where superscript, $(s)$, denotes the $s$th order derivative.
The proof and detailed discussion 
(including a comparison with the truncation errors of 
Shannon's sampling theorem) are given in Ref. \cite{qian} 
and are beyond the scope of this paper.

This theorem provides a guide to 
the choice of $M$, $\sigma$ and $\Delta$.
For example, in the case of interpolation ($s=0$), if the 
$L_2$ norm error is set to $10^{-\eta}$ ($\eta>0$), 
the following relations can be deduced from Eq. (\ref{eq30})
\begin{eqnarray}
r(\pi-B\Delta)>\sqrt{4.61\eta},
\end{eqnarray}
and
\begin{eqnarray}
{M\over r}>\sqrt{4.61\eta},
\end{eqnarray}
where $r=\sigma/\Delta$ (The choice of $\sigma$ is always 
proportional to 
$\Delta$ so that the width of the Gaussian envelope
varies with the central frequency). 
The first inequality states that for a given grid 
size $\Delta$, a large $r$ is required 
for approximating high frequency 
component of an $L^2$ function.
The second inequality indicates that 
if one chooses the ratio $r=3$, then the 
half bandwidth $M\sim30$ can be used to ensure the highest 
accuracy in a double precision computation ($\eta=15$). 
However, for lower accuracy requirement, a much 
smaller half bandwidth can be used.
In general, the value of $r$ is proportional to $M$.
The use of $M$ values is determined by the accuracy
requirement. 
This theoretical estimation is in excellent agreement with 
a previous numerical test\cite{weicpl}.

\section{Illustrative calculations}

In this section, we illustrate the use of the present 
approach for solving the Fokker-Planck equation 
and the incompressible Euler equation. 
Many DSC kernels are discussed 
in the previous work\cite{weijcp,weicpl} 
and they can be used as the DSC trial functions.
For simplicity, we focus on three DSC kernels,
a regularized Shannon's kernel (RSK), 
\begin{equation}\label{k1}
\phi^M_{{\pi\over\Delta},\sigma;k}(x)=
{\sin{\pi\over\Delta}(x-x_k)\over
{\pi\over\Delta}(x-x_k)}
\exp\left[ {-{(x-x_k)^2\over2\sigma^2}}\right],
\end{equation}     
a regularized Dirichlet kernel (RDK), 
\begin{equation}\label{k2}
\phi^M_{{\pi\over\Delta},\sigma;k}(x)=
{\sin\left[{\pi\over\Delta}(x-x_k)\right]\over
(2m+1)\sin\left[{\pi\over\Delta}{x-x_k\over2m+1}\right]}
\exp\left[ {-{(x-x_k)^2\over2\sigma^2}}\right],
\end{equation} 
and a regularized Lagrange kernel (RLK)
\begin{equation}\label{k3}
\phi^M_{{\pi\over\Delta},\sigma;k}(x)=
\prod_{i\neq k}^{2m}{x-x_i\over x_k-x_i}
\exp\left[ {-{(x-x_k)^2\over2\sigma^2}}\right],
\end{equation} 
for our numerical test.
Note that the resolution is given by 
$\alpha={\pi\over\Delta}$ which is the frequency bound in 
the Fourier representation. 
The goal of this section is to test
the present method 
for the solutions of the Fokker-Planck 
equation via time propagation and the incompressible 
Euler equation. 
For the numerical solution of the Fokker-Planck equation, 
we choose $\sigma=3.8 \Delta$ for 
the RSK and RDK, $\sigma=2.8 \Delta$ for the RLK, with 
$\pi/\Delta$ being the resolution. In fact, a wide range of 
$\sigma$ values can be used to deliver excellent 
results.  The half bandwidth, $M$,  
can be chosen to interplay between the local limit and the global
limit and is set to 40 in all calculations.
Finally, $m$ controls the order of the regularized Dirichlet and 
Lagrange kernels and is set to 40 in all calculations
(note that the selection of $m$ is independent of the 
grid used in the computation). It is noted that all of the
abovementioned DSC trial functions are of Schwartz class and are 
capable of auto-regularizing when used as integral kernels.
The fourth order explicit Runge-Kutta scheme is used 
for time discretization.
Details of these computations are 
described in the first three subsections. 
For treating the incompressible Euler 
equation, many other  DSC parameters are tested as
indicated in the last subsection.
Double precision is used in all calculations.

\subsection{The repulsive Wong process}

One of important stochastic systems is the 
repulsive Wong process\cite{Wong,Gisin,HonDes,PSMT}, 
given by
\begin{equation}\label{Wong1}
dx=2\gamma\tanh(x) dt+\sqrt{2}dF_t, 
\end{equation}
where $dF_t$ is the  Gaussian white noise which 
has the standard statistical properties
\begin{eqnarray}\label{Wong2}
<dF_t>=0
\end{eqnarray}
and 
\begin{eqnarray}\label{Wong2-2}
<dF_t, dF_\tau>=\delta(\mid t-\tau\mid).
\end{eqnarray}

The repulsive Wong process is Markovian 
due to the deriving Gaussian white noise term.
Its transition probability density is governed by the 
Fokker-Planck equation of the form\cite{Wong,Gisin,HonDes}
\begin{equation}\label{Wong3}
{\partial f(x,t)\over \partial t}
=-2\gamma  {\partial [\tanh(x)  
f(x,t)]\over\partial x}+ {\partial^2 f(x,t)\over\partial x^2},
\end{equation}
with the usual initial condition
\begin{equation}\label{Wong4}
f(x,0)=\delta(x-x_0),
\end{equation}
and the normalization
\begin{equation}\label{Wong5}
\int_{-\infty}^{\infty} f(x,t)dt=1.
\end{equation}
For $\gamma=1$, the solution\cite{Gisin,HonDes} of 
the Fokker-Planck equation (\ref{Wong3}) is analytically 
given by (for $x_0=0$) 
\begin{equation}\label{Wong6}
f(x,t)={1\over2\sqrt{4\pi t}}\left[e^{-{(x-x_-)^2 \over4t}}
+e^{-{(x-x_+)^2 \over4t}}\right],
\end{equation}
where $x_{\pm}=\pm2t$ are centers of two moving Gaussians.
Here the superposition of two Gaussians gives rise to
a monomodality-bimodality transition as time increases.
The Wong process is useful for illustrating  the 
connection between stochastic processes and quantum 
measurements. It is also useful for  distinguishing
spectrum differences between the Master equation 
and its Fokker-Planck equation approximations.

The accurate simulation of the Wong process is 
not a simple task because of the 
monomodality-bimodality transition.
Two Gaussian peaks centered at $x_{\pm}=\pm2t$ 
move apart as time increases.
The computational domain is to be sufficiently large 
in order to avoid boundary reflection (Otherwise, more 
complicated techniques, such as absorption boundaries, are 
to be implemented.). 
In the present computations, the resolution 
is chosen as ${\pi\over\Delta}=10\pi$.
The initial functions are approximated by a unit impulse 
function located at 0.
The equation (\ref{Wong3}) is 
integrated up to 10 time units with a time 
increment of 0.001. 
The errors for a wide
range of propagation times
are listed in TABLE I and  
are measured by error norms of $L_2$ and $L_\infty$.
It is seen  that the present unified 
approach is extremely accurate and 
reaches machine precision. All of the DSC kernels behave 
very similar to each other and provide the same level of 
accuracy and speed of convergence. In fact, other 
DSC kernels, such as regularized 
modified Dirichlet kernel, provide similar results.
The results of the RSK and RDK are slightly more 
accurate than those of the RLK. 
It is evident that the present unified method, 
in associated with the DSC trial functions,
is capable of delivering extremely high accuracy and 
numerical stability for the Wong process.
To our knowledge,
the DSC solution for this system is the best to the date.

\subsection{The Black-Scholes equation}

The Fokker-Planck equation and stochastic analysis have 
interesting applications  in mathematical 
modeling of financial market option pricing.
Consider a writer of a European call option on a stock, 
he is exposed to the risk of unlimited liability if the stock 
price rises acutely above the strike price.
To protect his short position in the option, he should consider 
purchasing certain amount of stock so that the loss in the short
position in the option is offset by the long position in the stock. 
In this way, he is adopting a hedging procedure. A hedge position
combines an option with its underlying 
asset so as to achieve the goal that either the stock protects the 
option against loss or the option  protects the stock against loss.
This risk-monitoring strategy has been commonly used by 
practitioners in financial markets. 
The most well-known stochastic model for the equilibrium condition 
between the expected return on the option, 
the expected return on the stock and the riskless interest rate is the 
Black-Scholes equation\cite{BlaSch}
\begin{equation}\label{BS}
{\partial c\over \partial t}={\nu^2\over2}
S^2 {\partial^2 c\over \partial S^2}+
rS {\partial c\over \partial S}-rc,
\end{equation} 
where   $S$ is the asset price which undergoes 
geometric Brownian motion, 
$c(S,t)$ the call price,
$\nu$ the volatility and $r$ the constant riskless interest rate.
Black-Scholes equation is a fundamental 
equation in finance and economics and is also an excellent 
example application of  stochastic analysis. 
By a simple transformation 
\begin{eqnarray}
x=\ln S,
\end{eqnarray}
and 
\begin{eqnarray}
f(x,t)=e^{rt}c(x,t),
\end{eqnarray}
 the Black-Scholes
equation is transformed into the Fokker-Planck equation of 
the standard form
\begin{equation}\label{BS2}
{\partial f\over \partial t}
= \left(r-{\nu^2\over2} \right)
{\partial f\over \partial x}
 + {\nu^2\over2}{\partial^2 f\over \partial x^2}.
\end{equation}

The numerical simulation of the Black-Scholes equation and 
its generalized versions is an important issue in financial
analysis and computational finance 
community\cite{BBG,BreSch,Tian,CRR}. 
Essentially, all existing numerical methods are 
tested for potential usefulness in estimating the 
option derivatives because both computational accuracy 
and efficiency are very important to option modeling
and risk estimation.
In the present time-dependent 
approach, the resolution is set to
${\pi\over\Delta}=2\pi$ and the time increment is chosen as 0.01.
For simplicity, we choose ${\nu^2\over2}=0.5$
and $r=0.7$ in our calculations.  
We chosen our initial  distribution 
as a unit impulse function located at $x=0$, which is 
a poor approximation to the true delta distribution.
Obviously, had one started with a smooth initial function,
or used a denser grid, one would have obtained much higher accuracy 
at earlier times as well. We have verified this computationally, but 
these results are not presented. 
Both $L_2$ and $L_{\infty}$ error analyses are used 
to evaluate the quality of the 
DSC approach, the results of which are listed in TABLE II. 
To our knowledge, the present time-dependent DSC approach provides  
the most accurate numerical results yet obtained for the 
Black-Scholes equation.

As in the first example, three DSC kernels provide extremely 
similar results in solving the Black-Scholes equation. 
This is not an isolated coincidence.
In fact, we can come up a number 
of other DSC kernels with all of their results being very similar
to those of the present three kernels.

\subsection{A nonlinear stochastic model}

To illustrate the accuracy and robustness of the present 
approach further, 
we choose the following nonlinear stochastic model 
\begin{equation}\label{Solve}
{\partial f(x,t)\over \partial t}={\partial [(\omega x +\theta
<x(t)>)f(x,t)]\over\partial x}+ 
D{\partial^2 f(x,t)\over\partial x^2},
\end{equation}
where $<x(t)>$ is the first moment of the distribution function
\begin{equation}\label{Solve1}
<x(t)>=\int^{\infty}_{-\infty}xf(x,t)dx,
\end{equation}
and $\omega, \theta$ and $D$ are constant. 
The initial probability distribution is also given by
\begin{equation}\label{Solve2}
f(x,0)=\delta(x-x_0).
\end{equation}
Equation (\ref{Solve}) is a true nonlinear stochastic
model since the instantaneous
position average depends on the distribution function.
This is one of few analytically soluble nonlinear systems
which are very valuable for testing new numerical
approaches. For example, Drozdov and Morillo have recently 
employed this system to test their K-point Stirling 
interpolation formula finite difference method\cite{DroMor}. 
The exact solution to Eq. (\ref{Solve})  is 
\begin{equation}\label{Solve3}
f(x,t)={1\over\sqrt{2\pi\nu(t)}}\exp\left[-{(x-<x(t)>)^2\over2\nu(t)}
\right],
\end{equation}
where $<x(t)>$ and $\nu(t)$ are analytically given by
\begin{equation}\label{Solve4}
<x(t)>=x_0{\rm e}^{-(\omega+\theta)t }   
\end{equation}
and 
\begin{equation}\label{Solve5}
\nu(t)={D\over\omega }\left(1-{\rm e}^{-2\omega t }\right),  
\end{equation}
respectively. Obviously, $\nu(t)$ is the theoretical value 
of the second moment $M_2(t)$
\begin{equation}\label{Solve6}
M_2(t)=<x^2(t)> - <x(t)>^2,
\end{equation}
which can also be used as a measure 
of computational accuracy.

In the present computations, the resolution 
is chosen as ${\pi\over\Delta}={239\over20}\pi$.
The time increment is taken as $\Delta t=0.005$.
In this example, the errors are measured 
by error norms of $L_1$ and $L_\infty$
from which all other error norms, such as the $L_2$ error norm,
can be interpolated. 
The $L_1$ and $L_{\infty}$ errors are listed in TABLE III, 
for $D=0.1, \omega=1, \theta=2$ and $x_0=2.0422$. 
The initial accuracy of computations is hindered 
by the poor approximation of the impulse function to the
Dirac delta function. 
However, the auto-regularization property of the Schwartz 
class trial functions enables the numerical integration to
stabilize at smooth solution and eventually
reach the machine precision at a slightly late time.

\subsection{The Euler equation}

All cases considered in the last three subsections are 
of strong parabolic type with a solution which becomes more 
and more flat and smooth as time increases. In this subsection,
we consider an additional problem, the incompressible 
Euler equation, to confirm that the results obtained 
for the Fokker-Planck equation are not due to the parabolic nature. 
We also use this example to demonstrate the 
inter-connection between the collocation and the finite 
difference, and between the local and the global.
It is hoped that this additional example helps to
build  confidence for using the DSC approach for 
treating more difficult problems.
Conceptually and numerically, it is convenient to describe 
the incompressible Euler equation from the point of view of 
the incompressible Navier-Stokes equation
\begin{eqnarray}\label{NS0}
&&{\partial {\bf v} \over\partial t}+{\bf v \cdot \nabla v}
=\nabla p +{1\over {\rm Re}}\nabla^2{\bf v},
\\ &&{\bf \nabla \cdot v} = 0,
\end{eqnarray}
where ${\bf v}$ is the velocity field vector,
$p$ is the pressure field and Re is the Reynolds number. 
The Euler equation is attained by setting Re$ =\infty$. 
Finding a general solution to the Euler equation is 
not an easy job. In the present study, we consider a 
solution domain of $[0,2\pi]\times [0,2\pi]$ with
periodic boundary conditions. Under such a constraint,
the Navier-Stokes equation (\ref{NS0}) exists an exact 
solution 
\begin{eqnarray}\label{NS6}
&&u(x,y,t)=-\cos(x)\sin(y)e^{-{2t\over {\rm Re}}} \nonumber\\
&&v(x,y,t)=\sin(x)\cos(y)e^{-{2t\over {\rm Re}}} \nonumber\\
&&p(x,y,t)=-{1\over4}[\cos(2x)+\cos(2y)]e^{-{4t\over {\rm Re}}},
\end{eqnarray}
where $(u, v)$ are the velocity components in the $x$-direction and
$y$-direction, respectively. Note that, for the Euler equation, 
the solution (\ref{NS6}) does not decay with time. 

In the present study, we use
a standard approach for treating the incompressible Navier-Stokes 
equation, i.e. deriving a Poisson equation for the pressure 
from the incompressible condition. 
The velocity fields are iterated by using  
the implicit Euler scheme. 
At time $t_{n+1}$, there are two coupled equations for 
the velocity fields 
\begin{eqnarray}\label{NS2}
&&\left({1\over {\rm Re}}\nabla^2-{1\over \Delta t}\right)u^{n+1}
=p^{n+1/2}_x+ S^n_x \\
&&\left({1\over {\rm Re}}\nabla^2-{1\over \Delta t}\right)v^{n+1}
=p^{n+1/2}_y+ S^n_y, \label{NS2-2}
\end{eqnarray}
and a Poisson equation for the pressure
\begin{eqnarray}\label{NS3}
&&\nabla^2p^{n+1/2}=S^n_p. 
\end{eqnarray}
Here $S^n_x, S^n_y$ and  $S^n_p$ are given by
\begin{eqnarray}\label{NS4}
&&S^n_x=-{u^{n}\over\Delta t}+(u^nu^n_x+v^nu^n_y) \nonumber\\
&&S^n_y=-{v^{n}\over\Delta t}+(u^nv^n_x+v^nv^n_y) \nonumber\\
&&S^n_p={1\over\Delta t}(u^n_x+v^n_y)-(u^n_x)^2-(v^n_y)^2-2u^n_yv^n_x. 
\end{eqnarray}
At each time $t_{n+1}$, the pressure field $p^{n+1/2}$ is solved   
according Eq. (\ref{NS3}) from the known 
velocity field vector $(u^n, v^n)$. 
The velocity field vector $(u^{n+1}, v^{n+1})$ is then 
updated according to Eqs. (\ref{NS2}) and (\ref{NS2-2}).
These linear algebraic equations are solved by using a standard
(LU decomposition) solver.

The derivatives in Eqs. (\ref{NS2}-\ref{NS3}) are 
computed by using the generalized finite difference 
scheme, Eq. (\ref{eq10}), and the required finite difference
weights are given by Eq. (\ref{eq11}). 
The involved trial functions, $\phi^M_{\alpha,\sigma}$, are given by 
the regularized Shannon's kernel (RSK) [Eq. (\ref{k1})], 
the regularized Dirichlet kernel (RDK) [Eq. (\ref{k2})], 
and the regularized Lagrange kernel (RLK) [Eq. (\ref{k3})].
Here $m=40$ is used for both RDK and RLK.
We choose a small time increment ($\Delta t=0.001$) so that 
the main error is caused by the spatial discretization. 
The number of grid points in each dimension is  
chosen as $N=4, 8, 16$  and 32 in various test calculations. 
The $\alpha$ value is specified as 
$\alpha={\pi\over\Delta}={\pi\over{2\pi\over N-1}}={N-1\over2}$.
For a given $N$, the matrix half bandwidth, $M$, can be chosen 
as $M\leq N$. In particular, if $M=N$, the approach has a 
global (full) computational matrix. For all $M<N$, the 
matrix is banded. In the present DSC approach, the 
connection between the global and the local can be easily
achieved by selecting an $M$ value for a given $N$. 
In particular, if $M\ll N$, the DSC approach
behaves truly like a finite difference scheme.  
To achieve optimal (or near optimal) accuracy, 
the $\sigma$ is chosen in proportional to $M$ and $\Delta$. 
When $M=32,16,8,4,2$ and 1, ${\sigma\over\Delta}$ are 
chosen as 3.2, 2.5, 1.8, 1.2, 0.9 and 0.6 
for both RSK and RDK,  
and 2.8, 2.0, 1.6 1.0 0.8 and 0.6 for RLK.
We compute the $L_2$ and $L_\infty$ errors of $u$
for a number of combinations of $N$ and $M$ and the results 
are listed  in TABLE IV for 4 different 
times ($t=0.5, 1.0, 1.5, 2.0$). 
A good consistent in accuracy among solutions at 
different times (or equivalently, over 2000 iterations) 
is observed. 
The DSC results are quite accurate when $N=M=4$
and are of machine precision when $N=M=32$. 
It is interesting to note that for fixed $M=4$, 
the results of $N=32$ and $N=4$ differ little in accuracy. 
We also checked the DSC-finite difference
approximation at the tridiagonal matrix 
limit ($M=1$) and the result is very good for 
$N=4$ (i.e. a total of 4 interior points in a 
$(2\pi)^2$ box).

\section{Discussion}

The main purpose of this paper is to discuss
the unified features of the discrete singular convolution 
(DSC) algorithm\cite{weijcp}.
It is found that the implementations of the DSC 
algorithm into a number of computational methods 
can be deduced from a single starting point,
the method of weighted residual.
This chain of deduction provides a unified  
approach for solving the Fokker-Planck equation
and other differential equations in general.
Some of these deduction relations are novel to our 
knowledge.

We demonstrate that 
by adjusting the support of the DSC trial 
functions, the DSC algorithm can be 
easily implemented either as a 
local method or as a global method.
For this reason, the present DSC approach has
 global method's accuracy while maintains  local 
method's flexibility for handling complex boundary
and geometry. In fact, the solution of the Fokker-Planck 
equation of a previous paper\cite{weijcp} was obtained by 
using the global limit. Whereas, in the present 
computations, a local approximation is used for all 
Fokker-Planck problems. A comparison between global and
local DSC treatments is given in solving the Euler 
equation.

We also show that the DSC implementations of
Galerkin and collocation are
computationally equivalent, i. e. 
the collocation, Eq. (\ref{eq6}) can be deduced   
from the Galerkin, Eq. (\ref{eq4}) because of the 
choice of the DSC trial functions.
Galerkin methods have a profound influence to the theory
of approximations. Both spectral methods and finite 
element methods are often formulated in the framework of 
the Galerkin approach. There has been a great deal of 
argument about advantage and disadvantage of the Galerkin 
in comparison to many other methods. The present DSC approach 
might provide a unified framework for the discussion of 
these  methods.

The present Galerkin-induced collocation scheme 
provides a nature base for the realization of 
finite difference methods. 
High order finite difference 
is not a new idea in numerical 
approximations\cite{Fornberg2}. 
However, the mathematical constructions of high 
order finite difference 
schemes often become too cumbersome to 
use in  practical 
applications as the order increases. 
The present DSC approach provides a
simple, systematic algorithm for the  generation of  
finite difference schemes of an arbitrary order.
The implementation of this finite difference is
demonstrated in solving the Euler equation with 
a number of different matrix bandwidths.

Recently, wavelet theory and techniques 
have had great success in signal processing, 
data compression, and telecommunication.
Two most important features of the wavelet theory
are multiresolution analysis and time-frequency localization.
Their potential applications in solving partial differential
equations  have been extensively 
explored\cite{DahKun,SchWyl,QiaWei,JawSwe,VPS,BeyKei}
in hope to come up with unified approaches for numerical
approximations. However, 
before wavelet approaches can be of  
practical use, a number of technical difficulties are to be 
overcome. In our view, the first difficulty is 
the implementation of boundary conditions 
in a multiresolution setting. 
The second difficulty is the requirement of  
sufficiently high wavelet regularity to provide
sufficiently weak solutions.
Moreover, there is a lack of general
and systematic  numerical algorithms 
for incorporating wavelets in an efficient manner.
Nevertheless, the wavelet multiresolution analysis
still has great potential for developing 
adaptive grid and multigrid algorithms. 
The present DSC algorithm is closely
related to the wavelet theory\cite{weijcp,weicpl}.
In fact, the DSC kernels have a feature in common
with wavelets in terms of time-frequency 
(position-momentum) localization. 
However, unlike in a wavelet algorithm, 
multiresolution analysis is feasible but 
it is not required in the DSC algorithm.

In contrast to our earlier work 
dealing with the application of the DSC approach to 
the Fokker-Planck equation via an eigenfunction 
expansion\cite{weijcp}, 
we have explored in this paper a DSC-based 
time-propagation approach for solving 
the Fokker-Planck equation.
Three typical  DSC kernels, the regularized Shannon's 
kernel (RSK), the regularized Dirichlet 
kernel (RDK), and the regularized Lagrange 
kernel (RLK), are used as trial functions 
in the framework of the present  method.
Four benchmark examples are chosen to demonstrate 
the usefulness and to test the accuracy of 
the present DSC approach.
The first example is the repulsive Wong process.
This is used for objectively testing the ability of 
handling the monomodality-bimodality transition.
The Wong process requires a large computational 
domain to ensure that the boundary reflection of 
density flux has little influence in a highly 
accurate computation.
By using reasonable resolution, regularization and 
a quite large time increment, the present 
approach performs very well in 
characterizing the transition. 
In fact, the present unified 
approach delivers machine accuracy at an early time.

The Black-Scholes equation of option pricing
was chosen as the second numerical example. 
This financial equation can be regarded as a 
reaction-diffusion 
equation, although, its derivation was entirely based 
stochastic analysis. By using a simple transformation,
the Black-Scholes equation is converted into the standard 
form of the Fokker-Planck equation which admits an analytical
solution. 
The present numerical results for the Black-Scholes equation
are obtained by using three 
different DSC kernels with a reasonable resolution 
and relatively large time mesh. 
The extremely high accuracy in the present calculation 
indicates that the DSC-based unified algorithm is a 
valuable potential 
approach for various  option pricing simulations.

The third example treated is a nonlinear stochastic 
model. The effective potential of the corresponding 
Fokker-Planck equation is time depended through the
first order moment of the transition probability density.
Despite of the  nonlinearity and
poor approximation of the initial density distribution,
the numerical solutions quickly settle to a smooth, stable and 
correct distribution after a few iterations. 
This is due to the fact that the DSC trial functions are chosen as 
Schwartz class functions
and they are capable of auto-regularizing when used as 
integration kernels.
Our results are of machine precision at a relatively late time.
To our knowledge, this is the best numerical 
solution to this nonlinear 
Fokker-Planck equation to the date. 
These illustrative calculations indicate that the present 
unified approach is extremely accurate, efficient and robust for
numerical simulations of stochastic systems.

A common feature in the abovementioned Fokker-Planck equation
is that the equation is of strong parabolic type and the 
solution decays as time increases. Therefore, it is necessary
to employ an additional example to   
validate the present DSC algorithm  further  
for handling more complicated partial differential
equations. To this end, we choose the incompressible 
Euler equation with its velocity field equations   
being of strong hyperbolic type and a derived equation for the
pressure being of elliptic type.  
A standard implicit Euler scheme is used for the time
discretization and at each time $t_{n+1}$, linear algebraic 
equations are constructed by using the 
collocation method. In the present approach, 
carrying out differentiations in the collocation 
is equivalent to implementing the finite difference 
weights computed from the DSC trial functions.
We test the DSC algorithm by using 
4, 8, 16 and 32 grid points in each dimension in 
association with many different half matrix bandwidths 
($M$= 4, 8, 16 and 32). As expected, the DSC 
algorithm achieves its highest accuracy at 
the global limit ($M=N$) for each given $N$. 
The machine precision is reached when $N=M=32$. 
Very good results are also obtained for many banded 
matrix calculations.
We believe that the feature of being 
able to provide both global and local 
approximations in one formulation is of 
practical importance for large scale computations.

Although this paper emphasizes the connection 
of a few computational methods 
and the unified features of the DSC approach, 
it claims neither that all computational methods are the same,
nor that the DSC algorithm engulfs all methods.
For example, it is still not clear whether the DSC 
algorithm is applicable in adaptive and unstructured grids
(progress is made on a DSC-multigrid method).
The reader is urged to keep the distinction 
of various methods in mind and maintain a
perspective.

\centerline{\bf Acknowledgment}

This work was supported in part by the National University of Singapore. 
The author is grateful to the referee for valuable 
comments and suggestions.


\newpage

\begin{table}
\caption{Errors for solving the repulsive Wong process}
\begin{center}
\begin{tabular}{c||c|c|c|c|c|c} 
   & \multicolumn{2}{c|}{RSK } & 
                    \multicolumn{2}{c|}{RDK} &
                    \multicolumn{2}{c}{RLK} 
\\ \cline{2-7}
 Time  & $L_2$ & $L_\infty$ 
       & $L_2$ & $L_\infty$ 
       & $L_2$ & $L_\infty$ \\ \hline
0.10 & 1.94(-09) & 2.32(-09) & 1.94(-09) & 2.32(-09) & 1.94(-09) & 2.32(-09)\\ 
0.25 & 4.67(-11) & 3.73(-11) & 4.67(-11) & 3.73(-11) & 4.67(-11) & 3.72(-11)\\ 
0.50 & 3.47(-12) & 2.05(-12) & 3.47(-12) & 2.05(-12) & 3.47(-12) & 2.03(-12)\\ 
0.75 & 8.76(-13) & 4.89(-13) & 8.76(-13) & 4.95(-13) & 8.83(-13) & 4.60(-13)\\ 
1.00 & 3.57(-13) & 1.83(-13) & 3.57(-13) & 1.91(-13) & 3.78(-13) & 1.91(-13)\\ 
2.00 & 6.29(-14) & 2.69(-14) & 5.92(-14) & 3.00(-14) & 2.18(-13) & 8.92(-14)\\ 
3.00 & 4.72(-14) & 2.04(-14) & 3.60(-14) & 1.63(-14) & 2.86(-13) & 9.76(-14)\\ 
4.00 & 5.23(-14) & 1.97(-14) & 3.76(-14) & 1.39(-14) & 3.53(-13) & 1.12(-13)\\ 
5.00 & 4.44(-14) & 1.19(-14) & 3.16(-14) & 1.12(-14) & 4.10(-13) & 1.21(-13)\\ 
6.00 & 5.87(-14) & 1.88(-14) & 5.23(-14) & 1.64(-14) & 4.68(-13) & 1.32(-13)\\ 
7.00 & 7.63(-14) & 2.38(-14) & 7.30(-14) & 2.16(-14) & 5.25(-13) & 1.43(-13)\\ 
8.00 & 9.27(-14) & 2.78(-14) & 9.08(-14) & 2.58(-14) & 5.80(-13) & 1.54(-13)\\ 
9.00 & 7.19(-14) & 1.99(-14) & 6.24(-14) & 1.79(-14) & 6.33(-13) & 1.63(-13)\\ 
10.0 & 6.98(-14) & 1.70(-14) & 5.25(-14) & 1.39(-14) & 6.87(-13) & 1.73(-13)\\ 
\end{tabular}
\end{center}
\end{table} 

\newpage

\begin{table}
\caption{Errors for the numerical solution of the Black-Scholes equation}
\begin{center}
\begin{tabular}{c||c|c|c|c|c|c} 
   & \multicolumn{2}{c|}{RSK } & 
                    \multicolumn{2}{c|}{RDK} &
                    \multicolumn{2}{c}{RLK} 
\\ \cline{2-7}
 Time  & $L_2$ & $L_\infty$ 
       & $L_2$ & $L_\infty$ 
       & $L_2$ & $L_\infty$ \\ \hline
1 & 1.85(-03)&  1.14(-03) & 1.83(-03)&  1.12(-03) & 2.18(-03)&  1.41(-03)\\
2 & 1.19(-04)&  8.55(-05) & 1.18(-04)&  8.45(-05) & 1.42(-04)&  1.06(-04)\\
3 & 4.57(-06)&  2.89(-06) & 4.48(-06)&  2.84(-06) & 6.44(-06)&  4.29(-06)\\ 
4 & 2.53(-07)&  1.68(-07) & 2.46(-07)&  1.63(-07) & 4.23(-07)&  2.84(-07)\\ 
5 & 1.81(-08)&  1.15(-08) & 1.75(-08)&  1.10(-08) & 3.66(-08)&  2.43(-08)\\ 
6 & 1.63(-09)&  1.09(-09) & 1.56(-09)&  1.04(-09) & 4.02(-09)&  2.74(-09)\\ 
7 & 1.81(-10)&  1.26(-10) & 1.71(-10)&  1.19(-10) & 5.48(-10)&  3.92(-10)\\ 
8 & 2.41(-11)&  1.59(-11) & 2.25(-11)&  1.48(-11) & 9.03(-11)&  6.05(-11)\\ 
9 & 3.83(-12)&  2.55(-12) & 3.54(-12)&  2.37(-12) & 1.76(-11)&  1.15(-11)\\ 
10 & 7.82(-13)&  6.25(-13)& 7.30(-13)&  5.85(-13) & 4.01(-12)&  2.84(-12)\\
20 & 4.86(-14)&  2.70(-14)& 4.91(-14)&  2.78(-14) & 4.84(-14)&  2.74(-14)\\
\end{tabular}
\end{center}
\end{table}

\newpage

\begin{table}
\caption{Errors for solving the nonlinear model}
\begin{center}
\begin{tabular}{c||c|c|c|c|c|c} 
   & \multicolumn{2}{c|}{RSK} & 
                    \multicolumn{2}{c|}{RDK} &
                    \multicolumn{2}{c}{RLK} 
\\ \cline{2-7}
 Time  & $L_1$ & $L_\infty$ 
       & $L_1$ & $L_\infty$ 
       & $L_1$ & $L_\infty$ \\ \hline
1 & 4.13(-01)&  3.09(-02) & 6.46(-01)&  4.71(-02) & 4.67(-02)&  3.98(-03)\\
2 & 4.66(-01)&  3.68(-02) & 7.41(-01)&  5.86(-02) & 2.48(-03)&  2.27(-04)\\
3 & 2.03(-01)&  1.68(-02) & 3.26(-01)&  2.68(-02) & 7.24(-04)&  6.51(-05)\\ 
4 & 4.88(-02)&  4.92(-03) & 7.97(-02)&  8.04(-03) & 1.23(-04)&  1.31(-05)\\ 
5 & 7.41(-03)&  7.84(-04) & 1.22(-02)&  1.29(-03) & 1.68(-05)&  1.82(-06)\\ 
6 & 1.02(-03)&  1.12(-04) & 1.68(-03)&  1.84(-04) & 2.26(-06)&  2.48(-07)\\ 
7 & 1.38(-04)&  1.52(-05) & 2.28(-04)&  2.50(-05) & 3.06(-07)&  3.36(-08)\\ 
8 & 1.87(-05)&  2.05(-06) & 3.09(-05)&  3.38(-06) & 4.14(-08)&  4.54(-09)\\ 
9 & 2.53(-06)&  2.77(-07) & 4.17(-06)&  4.58(-07) & 5.59(-09)&  6.13(-10)\\ 
10 & 3.42(-07)&  3.75(-08)& 5.65(-07)&  6.19(-08) & 7.56(-10)&  8.29(-11)\\
20 & 3.64(-14)&  3.11(-15)& 4.75(-14)&  4.88(-15) & 1.00(-13)&  1.51(-14)\\
\end{tabular}
\end{center}
\end{table}


\begin{table}
\caption{Errors for solving the Euler equation }
\begin{center}
\begin{tabular}{c|c|c||c|c|c|c|c|c} 
&  &  & \multicolumn{2}{c|}{RSK} & 
                    \multicolumn{2}{c|}{RDK} &
                    \multicolumn{2}{c}{RLK} 
\\ \cline{4-9}
$N$ &$M$ & Time   & $L_1$ & $L_\infty$ 
             & $L_1$ & $L_\infty$ 
             & $L_1$ & $L_\infty$ \\ \hline
4  & 1 & 0.5 &
3.16(-2)& 
6.12(-2)& 
3.15(-2)& 
6.12(-2)& 
3.09(-2)& 
6.01(-2)
\\
 &  & 1.0 &  
3.10(-2)& 
6.02(-3)& 
3.09(-2)& 
6.01(-3)& 
3.03(-2)& 
5.88(-3) 
\\
 &  & 1.5 &  
3.06(-2)& 
5.90(-3)& 
3.05(-2)& 
5.89(-3)& 
2.99(-2)& 
5.73(-3) 
\\
 &  & 2.0 &  
3.04(-2)&
5.77(-3)&
3.04(-2)&
5.76(-3)&
2.99(-2)&
5.57(-3)\\
\cline{2-9}
  & 2 & 0.5 & 
1.27(-2)& 
2.48(-3)& 
1.27(-2)& 
2.48(-3)& 
1.29(-2)& 
2.44(-3)\\
 &  & 1.0 & 
3.10(-2)& 
6.02(-3)& 
1.31(-2)& 
2.66(-3)& 
1.29(-2)& 
2.54(-3)\\
 &  & 1.5 & 
1.37(-2)& 
2.86(-3)& 
1.37(-2)& 
2.87(-3)& 
1.32(-2)& 
2.66(-3)\\
 &  & 2.0 &  
1.44(-2)& 
3.08(-3)& 
1.45(-2)& 
3.08(-3)& 
1.35(-2)& 
2.78(-3)\\
\cline{2-9}
  & 4 & 0.5 & 
9.33(-3)& 
1.70(-3)& 
9.32(-3)& 
1.69(-3)& 
9.34(-3)& 
1.70(-3)\\ 
  &  & 1.0 & 
9.43(-3)& 
1.79(-3)& 
9.42(-3)& 
1.79(-3)& 
9.44(-3)&
1.79(-3)\\ 
  &  & 1.5 & 
9.62(-3)& 
1.88(-3)& 
9.61(-3)& 
1.88(-3)& 
9.64(-3)&
1.88(-3)\\ 
  &  & 2.0 & 
9.92(-3)& 
1.96(-3)& 
9.91(-3)& 
1.95(-3)& 
9.93(-3)& 
1.96(-3)\\
\hline 
8 & 8 & 0.5 & 
1.30(-4) & 4.26(-5) &
1.33(-4)&  4.36(-5) &
1.24(-4)&  3.40(-5) \\ 
 & & 1.0 & 
1.52(-4)&  5.13(-5) &
1.54(-4)&   5.25(-5) &
1.54(-4)&   4.79(-5) \\ 
  & & 1.5 & 
1.82(-4) &  6.10(-5)  &
1.83(-4)&   6.26(-5)  &
1.92(-4) &  5.66(-5)  \\ 
  & & 2.0 & 
2.17(-4) &  7.13(-5) &
2.16(-4) &  7.32(-5) &
2.34(-4) &  6.72(-5) \\  
 \hline
16 & 16  & 0.5 &
6.30(-10)&  2.37(-10) &
6.75(-10)&  2.76(-10) &
1.23(-8) & 3.63(-9) \\ 
 & & 1.0 & 
6.76(-10)&   2.40(-10) &
6.82(-10) & 2.87(-10) &
1.56(-8) &  5.16(-9)  \\ 
  & & 1.5 & 
8.00(-10) &  2.65(-10)   &
7.57(-10)&   3.18(-10)  &
1.99(-8)  & 6.76(-9) \\ 
   & & 2.0 & 
9.68(-10)&   3.35(-10) &
8.81(-10) &  3.49(-10) &
2.48(-8)  & 8.53(-9)  \\ 
\hline
32 & 4 & 0.5 & 
5.25(-4)  &
2.10(-4) &
5.24(-4)  &
2.10(-4) &
2.37(-3)  &
7.14(-4)\\
  & & 1.0 & 
7.40(-4)  &
2.88(-4)  &
7.41(-4)  &
2.90(-4)  &
2.96(-3)  &
1.00(-3) \\
  & & 1.5 &
1.04(-3)  &
4.25(-4)  &
1.04(-3)  &
4.27(-4)  &
3.75(-3)  &
1.33(-3) \\
  & & 2.0 &
1.40(-3) &
5.87(-4) &
1.40(-3) &
5.89(-4) &
4.62(-3) &
1.67(-3)\\
\cline{2-9}
  & 8 & 0.5 & 
1.78(-6)  &
7.32(-7)  &
1.93(-6)  &
7.95(-7)  &
9.50(-7)  &
3.05(-7) \\
 & & 1.0 &
2.41(-6)  &
1.06(-6)  &
2.62(-6)  &
1.15(-6)  &
1.24(-6)  &
5.22(-7) \\
 & & 1.5 &
3.23(-6)  &
1.39(-6)  &
3.51(-6)  &
1.51(-6)  &
1.64(-6)  &
7.73(-7) \\
 & & 2.0 &
4.17(-6)  &
1.79(-6)  &
4.52(-6)  &
1.92(-6)  &
2.09(-6)  &
1.03(-6) \\
\cline{2-9}
 & 16 & 0.5 & 
6.95(-11) &
2.93(-11) &
9.52(-11) &
4.00(-11) &
1.90(-10) &
5.73(-11)\\
 &  & 1.0 & 
7.48(-11) &
3.23(-11) &
1.03(-10) &
4.42(-11) &
2.40(-10) &
8.15(-11) \\
  & & 1.5 &
8.06(-11) &
3.51(-11) &
1.11(-10) &
4.81(-11) &
3.04(-10) &
1.08(-10) \\
  & & 2.0 &
8.67(-11) &
3.79(-11) &
1.19(-10) &
5.19(-11) &
3.76(-10) &
1.36(-10)\\
\cline{2-9}
  &32  & 0.5 &
1.02 (-14) &  
6.99(-15) & 
1.36(-14)&   
1.21(-14) &
1.10(-14)&   
7.88(-15) \\
 & & 1.0 &
2.03(-14) &  
1.45(-14) &
2.51(-14) &  
1.80(-14) &
2.22(-14) &  
1.54(-14) \\ 
 & & 1.5 & 
2.98(-14) &  
1.88(-14) &
3.68(-14) &  
2.74(-14) &
3.54(-14) &  
2.22(-14) \\ 
  & & 2.0 &
4.05(-14) &  
2.31(-14) &
5.04(-14) &  
3.13(-14) &
4.80(-14) &  
3.12(-14)\\ 
\end{tabular}
\end{center}
\end{table}

\end{document}